\documentclass[runningheads]{llncs}
\usepackage[T1]{fontenc}
\usepackage{amsmath}
\usepackage{bigints}
\usepackage{tikz}
\usepackage{tikzscale}
\usepackage{pgfplots}
\usepackage{graphicx}
\usepackage{wrapfig}
\usepackage{comment}

\colorlet{myblue}{blue!70!black}
\colorlet{mylightblue}{blue!10}
\colorlet{branch}{blue!30!black}
\colorlet{evolcol}{blue!50!black}
\colorlet{natalcol}{red!50!white!60!black}
\colorlet{hormcol}{orange!70!black}
\colorlet{ethcol}{yellow!80!black}
\colorlet{stimcol}{red!80!black}
\colorlet{neurcol}{blue!80!black}

\usetikzlibrary{shapes,arrows,intersections}
\usetikzlibrary{fit,calc,matrix,decorations.markings,decorations.pathreplacing,decorations.text}

\begin{document}
\title{A theoretical framework for flow-compatible reconstruction of heart motion}
\titlerunning{Flow-compatible reconstruction of heart motion}
%

\author{
    Francesco Capuano\inst{1} \and
    Yue-Hin Loke\inst{2} \and
    Ibrahim Yildiran\inst{1} \and
    Laura J. Olivieri\inst{3} \and
    Elias Balaras\inst{1}
}

\authorrunning{F. Capuano et al.}

\institute{
    Department of Fluid Mechanics, Universitat Politècnica de Catalunya $\cdot$ BarcelonaTech, Barcelona, Spain \email{francesco.capuano@upc.edu} \and
    Children’s National Hospital, Washington DC 20010 \and
    Children's Hospital, Pittsburgh, PA 15224 \and
    Department of Mechanical and Aerospace Engineering, George Washington University, Washington, DC 20052
}

\maketitle{}          

\begin{abstract}
Accurate three-dimensional (3D) reconstruction of cardiac chamber motion from time-resolved medical imaging modalities is of growing interest in both the clinical and biomechanical fields. 
Despite recent advancement, the cardiac motion reconstruction process remains complex and prone to uncertainties. Moreover, traditional assessments often focus on static comparisons, lacking assurances of dynamic consistency and physical relevance. This work introduces a novel paradigm of flow-compatible motion reconstruction, integrating anatomical imaging with flow data to ensure adherence to fundamental physical principles, such as mass and momentum conservation. The approach is demonstrated in the context of right ventricular motion, utilizing diffeomorphic mappings and multi-slice MRI to achieve dynamically consistent and physically robust reconstructions. Results show that enforcing flow compatibility within the reconstruction process is feasible and enhances the physical realism of the resulting kinematics.

\keywords{Heart motion  \and Magnetic resonance imaging \and Right ventricle \and Congenital heart disease}
\end{abstract}

\section{Introduction}

The clinical and biomechanical communities are increasingly interested in obtaining accurate and reliable three-dimensional (3D) reconstructions of cardiac chamber motion from routine time-resolved medical imaging modalities, such as cardiac magnetic resonance imaging (MRI), computed tomography (CT), or three-dimensional echocardiography. Accurate 3D motion reconstructions have wide-ranging implications, including: (i) enhancing diagnostic tools by identifying subtle regional deformation abnormalities indicative of early-stage or subclinical disease~\cite{voigt20192}, (ii) isolating regions of interest in advanced imaging techniques, such as four-dimensional phase-contrast (4D Flow) MRI, to elucidate flow dynamics~\cite{soulat20204d}, or (iii) providing high-fidelity boundary conditions for image-driven biomechanical or electro-mechanical simulations~\cite{menon2024cardiovascular}.

Reconstructing a time- and space-resolved representation of cardiac motion poses significant challenges, requiring a pipeline of sophisticated operations. These typically include segmentation of cardiac regions of interest, geometric reconstruction via the generation of triangulated surface models from segmented images, registration of the 3D models across the cardiac cycle, and temporal interpolation~\cite{mittal2016computational}. While recent advances in machine learning have partially alleviated the computational burden of segmentation and model registration~\cite{chen2020deep}, the overall process remains labor-intensive, computationally demanding, and sensitive to uncertainties introduced at various stages. Importantly, most existing methods assess results \textit{statically}, comparing reconstructed shapes to ground-truth images at discrete time frames using similarity metrics such as the Dice coefficient or Hausdorff distance~\cite{morales2019implementation}. Such evaluations offer limited insights into the motion’s \textit{dynamic} consistency, which is crucial to ensure the physiological and physical plausibility of the derived kinematics.

To address these limitations, this work introduces the novel paradigm of \textit{flow-compatible} motion reconstruction for the heart. Given the increasing availability of blood flow data (via, e.g., phase-contrast MRI or Doppler-based echocardiographic techniques) as part of routine cardiac imaging protocols, the proposed approach integrates flow data to ensure that reconstructed cardiac motion adheres to fundamental physical principles, such as global mass and momentum conservation. Specifically, it is conjectured that combining anatomical imaging with complementary flow data can: (i) provide a rigorous assessment of the dynamic quality of motion reconstructions, and (ii) enforce flow compatibility within the reconstruction pipeline itself. This integration minimizes uncertainty and enhances the physical fidelity of kinematic models by leveraging the coupling between shape (anatomical structure) and function (blood flow dynamics).

This paper introduces the theory of flow-compatible reconstruction and applies it to the challenging problem of right ventricle (RV) motion. RV motion poses unique difficulties due to its highly asymmetrical geometry, peculiar deformation pattern, and intricate interactions with blood flow. The work builds upon a recently developed methodology employing diffeomorphic mappings to reconstruct cardiac motion from multi-slice MRI. By ensuring compatibility with observed flow data, this approach represents a significant step toward generating dynamically consistent and physically robust cardiac motion reconstructions.

\section{Methods}

\subsection{Theoretical framework}

\begin{figure}
  \centering \includegraphics[width=.46\textwidth]{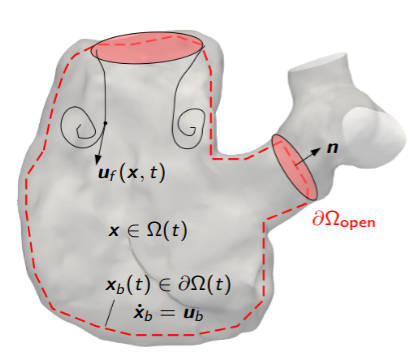}
  \caption{Sketch of a prototype heart chamber filled with blood flow and nomenclature used in this work. The red planes are indicative of the permeable (valve) surfaces.}
  \label{fig:sketch}
\end{figure}

A prototype heart chamber is shown schematically in Figure \ref{fig:sketch}, along with the nomenclature used in this work. The Eulerian blood flow velocity within the chamber is $\mathbf{u}_f(\mathbf{x},t)$, where $\mathbf{x} \in \Omega(t)$, with $\Omega$ the domain of interest. On the other hand, $\mathbf{x}_b(t)$ \textcolor{black}{and $\mathbf{u}_b(t) = \mathbf{\dot{x}}_b$ describe, respectively, the position and velocity} of the endocardium $\partial \Omega$, including both closed portions (walls) and open surfaces (valves). Reconstructing a high-fidelity and physically-consistent estimate of $\mathbf{x}_b(t)$ is the ultimate objective of this work. 

The endocardium velocity is linked to the blood flow velocity through integral relations that can be obtained applying the Reynolds transport theorem (RTT)
\begin{equation}
 \dfrac{\text{d}\textcolor{black}{P}^{\text{syst}}}{\text{d}t}  = \dfrac{\text{d}}{dt}\int\displaylimits_{\Omega(t)} \rho \textcolor{black}{p} \, \text{d}V + \int\displaylimits_{\partial \Omega(t)} \rho \textcolor{black}{p} \underbrace{\left(\mathbf{u}_f - \mathbf{u}_b \right)}_{\mathbf{u}_r} \cdot \mathbf{n} \, \text{d}S  \label{eq:RTT}
\end{equation}
to the moving and deforming control volume shown in Figure~\ref{fig:sketch}. In Eq.~\eqref{eq:RTT}, $\rho$ is blood density, $\mathbf{n}$ is the surface normal, \textcolor{black}{$P$ is a generic extensive property of the system, $p$ the corresponding intensive property and $P^{\text{syst}}$ is the extensive property of the material volume contained in $\Omega(t)$;} $\mathbf{u}_r$ is the relative velocity.

\textcolor{black}{Mass conservation can be analyzed setting $P^{\text{syst}} = m^{\text{syst}}$, where $m$ is the blood mass; in this case, $p = 1$.} Taking into account that $\rho$ is constant and splitting the control surface into open (permeable) and closed (impermeable) portions, the global mass conservation balance reads:
\begin{equation}
    \int\displaylimits_{\partial \Omega_\text{open}}  \mathbf{u}_f  \cdot \mathbf{n} \, \text{d}S 
       \stackrel{\text{def.}}{=} q(t) = -\int\displaylimits_{\partial \Omega_\text{closed}}\mathbf{u}_b \cdot \mathbf{n} \, \text{d}S \,, \label{eq:mass_balance}
\end{equation}
\textcolor{black}{where the \textit{flux} scalar quantity $q(t)$ has been defined, representing the volumetric blood flow rate across open surfaces.}

On the other hand, momentum conservation can be studied by setting $\textcolor{black}{P}^{\text{syst}} = (m\mathbf{u})^{\text{syst}}$ and therefore $\textcolor{black}{p} = \mathbf{u}_f$, yielding
\begin{equation*}
    \begin{split}     
       \dfrac{\text{d}(m\mathbf{u}_f)^{\text{syst}}}{\text{d}t} \stackrel{\text{def.}}{=} \mathbf{F}(t) &= \dfrac{\text{d}}{\text{d}t}\int\displaylimits_{\Omega(t)} \rho \mathbf{u}_f \, dV + \int\displaylimits_{\partial \Omega(t)} \rho \mathbf{u}_f \mathbf{u}_r \cdot \mathbf{n} \, \text{d}S = \\
        &\stackrel{\text{Leibniz}}{=} \rho  \int\displaylimits_{\Omega(t)} \dfrac{\partial  \mathbf{u}_f}{\partial t}  \, dV + \rho \int\displaylimits_{\partial \Omega(t)} \mathbf{u}_f \mathbf{u}_f \cdot \mathbf{n} \, \text{d}S .
    \end{split}
\end{equation*}
Using a recently derived vector identity~\cite{pedrizzetti2019computation}, the hemodynamic force $\mathbf{F}(t)$ exchanged between the blood and the endocardium walls can be ultimately expressed as a surface (rather than a volume) integral:
\begin{equation}
       \mathbf{F}(t) = \rho  \bigintsss\displaylimits_{\partial \Omega(t)} \left[ \mathbf{x}_b \left( \dfrac{\partial  \mathbf{u}_f}{\partial t} \cdot \mathbf{n} \right)   + \mathbf{u}_f \mathbf{u}_f \cdot \mathbf{n} \right] \, \text{d}S \label{eq:momentum_balance}
\end{equation}
Of note, in the r.h.s. of Eq.~\eqref{eq:momentum_balance} the flow velocity $\mathbf{u}_f$ is equal to the endocardium velocity $\mathbf{\dot{x}}_b$ on the closed surfaces, and can be related to $\mathbf{x}_b$ also on the permeable surfaces via mass conservation. Therefore, Eqs.~\eqref{eq:mass_balance} and \eqref{eq:momentum_balance} constitute two relationships between the boundary displacement field (and its derivatives) and the blood flow field. These relationships are exact at a continuous level, but are not necessarily satisfied when $\mathbf{x}_b$ and its derivatives are reconstructed from discrete data. The idea proposed in the paper is to use Eqs.~\eqref{eq:mass_balance} and \eqref{eq:momentum_balance} as physical constraints to (i) assess the  consistency of reconstructed motion dynamics, and (ii) enforce flow compatibility within the reconstruction pipeline itself.

\subsection{Flow-compatible motion reconstruction}

In a clinical setting, an estimate \textcolor{black}{of the flux and hemodynamic force vector \textit{in vivo} can be obtained through imaging techniques; the corresponding quantities are hereinafter denoted with a tilde. For instance, measurements} $\tilde{q}(t)$ of the flux at one or more locations (typically valve planes) are often available with \textit{reasonable} accuracy, e.g., $\pm 5\%$, using phase-contrast MRI~\cite{hansen2014method}. Conversely, estimating $\tilde{\mathbf{F}}(t)$, the hemodynamic force vector, requires knowledge of the complete blood flow field within the cardiac chamber of interest, which typically relies on 4D flow MRI; its uncertainties, however, remain to be fully assessed. The discrete anatomical information regarding the endocardium position, obtained via echo, MRI or CT, is hereinafter denoted as $\tilde{\mathbf{x}}_b$.

In the overwhelming majority of geometric and motion reconstruction approaches, the objective is to find $\mathbf{x}_b(t=t_i)$ to be \textit{as close as possible} (as measured by a proper distance function) to $\tilde{\mathbf{x}}_b(t_i)$. However, as shown in the previous section, the position field and its derivatives should also satisfy additional constraints, which are rewritten compactly as follows:
\begin{subequations}\label{eq:constraints}
\begin{align}       
    q(t) &= f_{\text{mass}} (\mathbf{\dot{x}}_b) \\
    \mathbf{F}(t) &= \mathbf{f}_{\text{mom}}(\mathbf{x}_b,\mathbf{\dot{x}}_b,\mathbf{\ddot{x}}_b)
\end{align}
\end{subequations}
Therefore, the \textit{flow-compatible} motion reconstruction process can be ultimately stated as a variational problem:
\begin{equation}
\begin{split}
        \mathbf{x}_b(t) = \: &\text{argmin}
        \Bigl[ 
        \gamma_1 \|  \tilde{\mathbf{x}}_b(t_i) - \mathbf{x}_b(t=t_i)\|_V +  \\
       &+  \gamma_2 \left| \tilde{q}(t)-f_{\text{mass}} \right| +
        \gamma_3 \| \tilde{\mathbf{F}}_b(t_i)-\mathbf{f}_{\text{mom}} \|_2
        \Bigr], \label{eq:variational}
\end{split}
\end{equation}
where $\gamma_i$ are weighing coefficients and $\| \cdot \|_V$ is a proper distance function. Evidently, $\gamma_2 = \gamma_3 = 0$ is the \textit{standard} reconstruction procedure which is exclusively based on anatomical information, while the addition of the extra terms constitutes the novel flow-compatible procedure. \textcolor{black}{It is important to observe that these extra penalization terms are global, i.e., they depend on the entire set of points $\mathbf{x}_b(t)$, reflecting the integral nature of the flow-compatible relations. The inclusion of further flow-based local constraints may be addressed in future work.}

\subsection{Multi-slice MRI reconstruction of RV motion}

The novel paradigm is illustrated building upon a recently developed pipeline
to reconstruct heart motion (and specifically right ventricular motion) from conventional MRI data, which is described in the following.

The methodology relies on: (i) a 3D end-diastolic anatomical dataset either involving 3D steady state free precession (3D SSFP) or magnetic resonance angiography (MRA); (ii) multi-slice cine-MRI imaging of the RV from both long-axis (LAX) and short-axis (SAX) views. The MRI parameters for acquisition vary with patient size, heart rate and lab standards, but should have optimized spatial/temporal resolution with reconstruction of 30 cardiac phases, allowing for diagnostic quality measurements of end systolic and end diastolic volumes. 

The detailed 3D end-diastolic model of the RV is used as the basis for the kinematics reconstruction process. Any segmentation method can be used for this step. In this work, models of the RV were created from MRA and 3D SSFP datasets, according to lab standards, using commercially available (Mimics; Materialise, Leuven, Belgium) or open-source (3D Slicer) software. The 3D model incorporated the three components of the RV including RV inflow, RV body/apex, and RV outflow tract (RVOT). The result of the segmentation is a triangulated surface mesh constituted by $N_p$ points with coordinates $\mathbf{x}_b(t=t_0)$. For convenience, it is assumed that $t_0$ corresponds to the end-diastolic frame, $t/T = 0$, where $T$ is the duration of the cardiac cycle.
For further analyses, the planes of the tricuspid/pulmonary valve annulus are also carefully delineated by tagging the triangular elements and corresponding vertices on the 3D model.


The multi-slice cine-MRI data is processed to capture the RV motion throughout the cardiac cycle. 
In this work, tissue tracking of the endocardium contour was performed using the commercial software QStrain v2.0 (Medis Medical Imaging Systems, Leiden, Netherlands), a semi-automated process that requires manual contour of the end-diastolic RV endocardial border to initiate so-called \textit{feature tracking} (FT). FT is a relatively mature technology that allows to recognize the same pattern (e.g., the endocardial border) within two images separated by a short time interval \cite{claus2015tissue}. The software provides the trajectory of $N_\text{SAX}^m = 48$ and $N_\text{LAX}^m = 49$ markers, placed along the contours of each slice. The resulting point clouds constitute the anatomical information available, and according to the nomenclature used in this work are indicated as $\tilde{\mathbf{x}}_b^\text{SAX}(t_i)$ and $\tilde{\mathbf{x}}_b^\text{LAX}(t_i)$, where $i=0, \dots, N_t-1$, where $N_t$ is the number of available timeframes (typically between 20 and 30 in standard MRI sequences). The vectors $\tilde{\mathbf{x}}_b^\text{SAX}(t_i)$ and $\tilde{\mathbf{x}}_b^\text{LAX}(t_i)$ are rearranged into vectors of sizes $N^m_\text{SAX} \times N_\text{SAX}$ and $N^m_\text{LAX} \times N_\text{LAX}$ respectively, where  $N_\text{SAX}$ (resp. $N_\text{LAX})$ is the number of slices in the short-axis (resp. longitudinal) planes. Full short-axis coverage of the RV is usually required (resulting $N_\text{SAX} \approx 8-9$); on the other hand, 3-chamber and 4-chamber slices are usually enough for a good reconstruction of the RV motion. Nonetheless, if additional slices are available in the longitudinal plane (or even in other planes), they can be tracked and included in the algorithm as well. 
Alternative options to track the endocardium contours (e.g., non-linear registration techniques, as opposed to vendor-dependent FT algorithms) will be explored in future work.



To capture the motion of the RV, the data obtained from FT analysis are used to compute mappings from the end-diastolic phase of the point clouds to each of the phases of the cardiac cycle. In particular, a family of transformations belonging to the large deformation diffeomorphic metric mapping (LDDMM) framework \cite{beg2005computing,durrleman2014morphometry} is employed. One of the main advantages of using the LDDMM approach is that it provides a smooth global mapping, it inherently preserves the topology of surfaces and avoids singularities or self-intersections of the surface mesh even for large deformations. In this framework, the time evolution of a generic set of points, $\mathbf{x}$,
is parametrized by a sparse combination of a discrete set of $N_c$ control points, $\mathbf{c}_p$, and corresponding momenta, $\boldsymbol{\alpha}_p$, as follows:
\begin{equation}
\dot{\mathbf{x}}(t) = \sum_{p=1}^{N_c} K
\left(\mathbf{x},\mathbf{c}_p(t) \right) \boldsymbol{\alpha}_p(t) \;,
\label{eq:points}
\end{equation}
where $K$ is a Gaussian kernel given by 
$K(x,y) = \exp{\left(-\|x-y \|^2/\sigma_v^2 \right)}$, and $\sigma_v$
is a parameter controlling the width of the kernel. Use of
Eq.~\eqref{eq:points} allows to compute a transformation
(\textit{registration}) between an initial \textit{reference} state
$\mathbf{x}^\text{ref}$ and a \textit{target} state
$\mathbf{x}^{\text{targ}}$, through specification of $\mathbf{c}_p(t)$
and $\boldsymbol{\alpha}_p(t)$. In turn, the control points and
momenta evolve according to similar evolution equations, such that
%
%
the dynamics of the entire system,
is completely determined by the
initial state $\mathbf{S}(0) = \{\mathbf{c}(0), \boldsymbol{\alpha}(0) \}$. 
Computing the transformation is an optimization process that consists in finding the state $\mathbf{S}(0)$ that minimizes the Euclidean distance between the target and the integrated reference states.

The idea is to deform the end-diastolic representation of the RV, $\mathbf{x}_b(t=t_0)$, according to the motion of the markers. This is achieved by computing a total of $2 \times N_t$ transformations, from the reference frames $\tilde{\mathbf{x}}_b^\text{SAX}(t_0)$ and $\tilde{\mathbf{x}}_b^\text{LAX}(t_0)$, to each of the corresponding instances $\tilde{\mathbf{x}}_b^\text{SAX}(t_n)$ and $\tilde{\mathbf{x}}_b^\text{LAX}(t_n)$. The two mapping fields are ultimately superimposed to obtain the desired three-dimensional motion. To this end, it is observed that the RV contraction/relaxation pattern occurs along three main components: (i) longitudinal shortening, with tricuspid annulus being pulled towards the apex, (ii) radial motion, with inward movement of the free wall, and (iii) anteroposterior shortening, by stretching of the free wall over the septum~\cite{kovacs2019right}. It is conjectured that the SAX cine stack captures radial and anteroposterior movement, while the LAX slices capture the longitudinal component. Therefore, the non-longitudinal component of the longitudinal markers is removed at each time $n = 1, \dots, N_t -1$:
\begin{equation}
\tilde{\mathbf{x}}_b^\text{LAX}(t_n) = \tilde{\mathbf{x}}_b^\text{LAX}(t_0) + \left[ \left( \tilde{\mathbf{x}}_b^\text{LAX}(t_n) - \tilde{\mathbf{x}}_b^\text{LAX}(t_0) \right) \cdot \hat{\mathbf{e}}_\text{a-b} \right] \hat{\mathbf{e}}_\text{a-b},  
\end{equation}
where $\hat{\mathbf{e}}_\text{a-b}$ is the unit vector normal to the SAX plane.

The computed diffeomorphisms are finally applied to the segmented 3D end-diastolic model of the RV, to obtain $N_t$ triangulated surface meshes with the same connectivity properties and with vertices correspondence.

\subsection{Implementation of flow-compatible reconstruction}

The flow-compatible paradigm of Eq. \eqref{eq:variational} can be implemented at various stages of the reconstruction procedure outlined above. It is found that it is computationally more efficient to implement the procedure as a \textit{correction} step, i.e., to first generate the sequence of 3D models of the RV using the algorithm described in the previous section and based only on anatomical information (i.e., using $\gamma_2 = \gamma_3 = 0$). Then, the 3D models are re-registered (corrected) using again LDDMM and solving the optimization problem in Eq. \eqref{eq:variational}, this time including the desired flow-related terms. Note that this approach can be applied to any set of registered 3D models of the heart chambers, regardless of the imaging modality and/or process used to generate them. \textcolor{black}{The optimization problem is solved using the Broyden–Fletcher–Goldfarb–Shanno (BFGS) quasi-Newton algorithm implemented within the \texttt{fminunc} function of MATLAB (Mathworks, USA).}

\section{Results and discussion}

The proposed algorithm was evaluated on a dataset of $N=12$ pediatric subjects, of which 6 healthy volunteers and 6 patients with repaired Tetralogy of Fallot (rTOF). 
All CMR studies were performed with a Siemens 1.5T scanner.
CMR data included cine imaging (long-axis and short-axis cine), MRA and 3D SSFP, two-dimensional phase contrast across the pulmonary valve 
and 4D flow. 

\begin{figure}
\centering
    \begin{tabular}{ccc}
  \begin{tikzpicture}
    \pgfplotsset{
        height=4cm, width=4cm,
        every tick label/.append style={font=\scriptsize}
    }
    \begin{axis}
    [
    ylabel={\footnotesize $\tilde{q}, f_{\text{mass}}$  [L/s]},
    y tick label style={
    /pgf/number format/.cd,
    precision=4,
    /tikz/.cd,
    },
    xmin=0, xmax=1, ymin = -1, ymax = 1,
    xlabel absolute, xlabel style={yshift=0.1cm},
    ]
    \addplot[black,   solid, line width=.25pt] table [x index={0},y index={1}] {./data/rtof_3_closed};
    \addplot[only marks,mark size = 1.5,red,opacity=0.4,
            mark=*,] table [x index={0},y index={1}] {./data/rtof_3_open};
    \node[above,black] at (axis cs:0.2,0.55){\scriptsize \textcolor{gray}{tof3}}; 
    \end{axis}
  \end{tikzpicture}
    & 
    \hspace{-0.35cm}
  \begin{tikzpicture}
    \pgfplotsset{
        height=4cm, width=4cm,
        every tick label/.append style={font=\scriptsize}
    }
    \begin{axis}
    [
    y tick label style={
    /pgf/number format/.cd,
    precision=4,
    /tikz/.cd,
    },
    xmin=0, xmax=1, ymin = -0.75, ymax = 0.75,
    ylabel absolute, ylabel style={yshift=-0.2cm},
    xlabel absolute, xlabel style={yshift=0.1cm},
    ]
    \addplot[black,   solid, line width=.25pt] table [x index={0},y index={1}] {./data/rtof_4_closed};
    \addplot[only marks,mark size = 1.5,red,opacity=0.4,
            mark=*,] table [x index={0},y index={1}] {./data/rtof_4_open};
    \node[above,black] at (axis cs:0.2,0.4){\scriptsize \textcolor{gray}{tof4}};
    \end{axis}    
  \end{tikzpicture}
      & 
    \hspace{-0.35cm}
  \begin{tikzpicture}
    \pgfplotsset{
        height=4cm, width=4cm,
        every tick label/.append style={font=\scriptsize}
    }
    \begin{axis}
    [
    y tick label style={
    /pgf/number format/.cd,
    precision=4,
    /tikz/.cd,
    },
    xmin=0, xmax=1, ymin = -0.75, ymax = 0.75,
    ylabel absolute, ylabel style={yshift=-0.2cm},
    xlabel absolute, xlabel style={yshift=0.1cm},
    ]
    \addplot[black,   solid, line width=.25pt] table [x index={0},y index={1}] {./data/rtof_5_closed};
    \addplot[only marks,mark size = 1.5,red,opacity=0.4,
            mark=*,] table [x index={0},y index={1}] {./data/rtof_5_open};
    \node[above,black] at (axis cs:0.2,0.4){\scriptsize \textcolor{gray}{tof5}};
    \end{axis}    
  \end{tikzpicture}
  \\[0.2cm]
  \begin{tikzpicture}
    \pgfplotsset{
        height=4cm, width=4cm,
        every tick label/.append style={font=\scriptsize}
    }
    \begin{axis}
    [
    xlabel={\footnotesize $t/T$},
    ylabel={\footnotesize $\tilde{q}, f_{\text{mass}}$  [L/s]},
    y tick label style={
    /pgf/number format/.cd,
    precision=4,
    /tikz/.cd,
    },
    xmin=0, xmax=1, ymin = -0.5, ymax = 0.5,
    xlabel absolute, xlabel style={yshift=0.1cm},
    legend entries={$\tilde{q}(t)$, $f_\text{mass}$},
    legend style={fill=none, draw=none, legend pos = south east,font=\scriptsize},   
    ]
    \addplot[only marks,mark size = 1.5,red,opacity=0.4,
            mark=*,] table [x index={0},y index={1}] {./data/normal_1_open};
    \addplot[black,   solid, line width=.25pt] table [x index={0},y index={1}] {./data/normal_1_closed};
    \node[above,black] at (axis cs:0.18,0.28){\scriptsize \textcolor{gray}{n1}}; 
    \end{axis}
  \end{tikzpicture}
 
&    
    \hspace{-0.35cm}
  \begin{tikzpicture}
    \pgfplotsset{
        height=4cm, width=4cm,
        every tick label/.append style={font=\scriptsize}
    }
    \begin{axis}
    [
    xlabel={\footnotesize $t/T$},
    y tick label style={
    /pgf/number format/.cd,
    precision=4,
    /tikz/.cd,
    },
    xmin=0, xmax=1, ymin = -0.5, ymax = 0.5,
    xlabel absolute, xlabel style={yshift=0.1cm},
    ]
    \addplot[black,   solid, line width=.25pt] table [x index={0},y index={1}] {./data/normal_2_closed};
    \addplot[only marks,mark size = 1.5,red,opacity=0.4,
            mark=*,] table [x index={0},y index={1}] {./data/normal_2_open};
    \node[above,black] at (axis cs:0.18,0.28){\scriptsize \textcolor{gray}{n2}}; 
    \end{axis}
  \end{tikzpicture}
    & 
    \hspace{-0.35cm}
    \begin{tikzpicture}
    \pgfplotsset{
        height=4cm, width=4cm,
        every tick label/.append style={font=\scriptsize}
    }
    \begin{axis}
    [
    xlabel={\footnotesize $t/T$},
    y tick label style={
    /pgf/number format/.cd,
    precision=4,
    /tikz/.cd,
    },
    xmin=0, xmax=1, ymin = -0.5, ymax = 0.5,
    xlabel absolute, xlabel style={yshift=0.1cm},
    ]
    \addplot[black,   solid, line width=.25pt] table [x index={0},y index={1}] {./data/normal_6_closed};
    \addplot[only marks,mark size = 1.5,red,opacity=0.4,
            mark=*,] table [x index={0},y index={1}] {./data/normal_6_open};
    \node[above,black] at (axis cs:0.18,0.28){\scriptsize \textcolor{gray}{n6}}; 
    \end{axis}
  \end{tikzpicture}

   \end{tabular}
   \caption{Mass conservation of RV motion reconstruction for selected normal subjects and patients. Flow compatibility is not enforced ($\gamma_2 = \gamma_3 = 0$). \label{fig:mass_rtof}}
\end{figure}
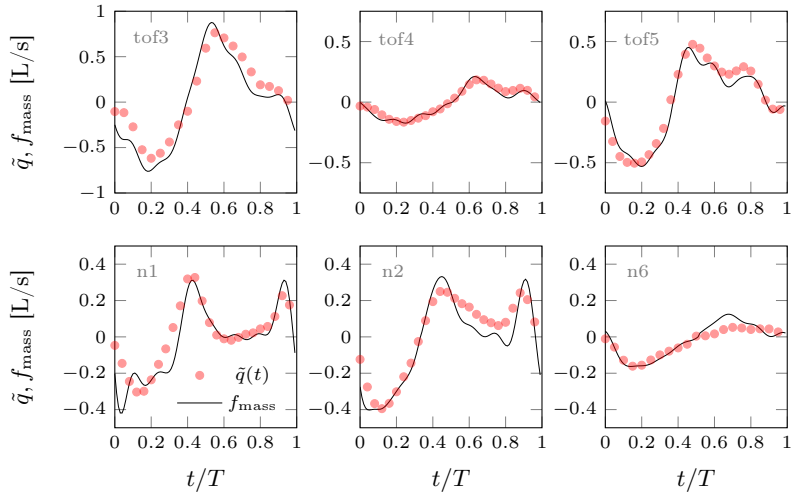

\begin{figure}
\centering
    \begin{tabular}{ccc}
  \begin{tikzpicture}
    \pgfplotsset{
        height=4cm, width=4cm,
        every tick label/.append style={font=\scriptsize}
    }
    \begin{axis}
    [
    ylabel={\footnotesize $\tilde{\mathbf{F}}, \boldsymbol{f}_{\text{mom}}$  [N/L]},
    y tick label style={
    /pgf/number format/.cd,
    precision=4,
    /tikz/.cd,
    },
    xmin=0, xmax=1, ymin = -2.5, ymax = 2.5,
    ylabel absolute, ylabel style={yshift=-0.2cm},
    ]
    \addplot[green, solid, line width=.25pt] table [x index={0},y index={1}] {./data/hdf_rtof_3_cine};
    \addplot[blue, solid, line width=.25pt] table [x index={0},y index={2}] {./data/hdf_rtof_3_cine};
    \addplot[brown, solid, line width=.25pt] table [x index={0},y index={3}] {./data/hdf_rtof_3_cine};
    \addplot[only marks,mark size = 1.5,green,opacity=0.4,
            mark=*,] table [x index={0},y index={1}] {./data/hdf_rtof_3_4d};
    \addplot[only marks,mark size = 1.5,blue,opacity=0.4,
            mark=*,] table [x index={0},y index={2}] {./data/hdf_rtof_3_4d};
    \addplot[only marks,mark size = 1.5,brown,opacity=0.4,
            mark=*,] table [x index={0},y index={3}] {./data/hdf_rtof_3_4d};

    \node[above,black] at (axis cs:0.82,-2.2){\scriptsize \textcolor{gray}{tof3}};                
    \end{axis}
  \end{tikzpicture}
 
&    
    \hspace{-0.35cm}
  \begin{tikzpicture}
    \pgfplotsset{
        height=4cm, width=4cm,
        every tick label/.append style={font=\scriptsize}
    }
    \begin{axis}
    [
    y tick label style={
    /pgf/number format/.cd,
    precision=4,
    /tikz/.cd,
    },
    xmin=0, xmax=1, ymin = -2, ymax = 2,
    ]
    \addplot[green, solid, line width=.25pt] table [x index={0},y index={1}] {./data/hdf_rtof_4_cine};
    \addplot[blue, solid, line width=.25pt] table [x index={0},y index={2}] {./data/hdf_rtof_4_cine};
    \addplot[brown, solid, line width=.25pt] table [x index={0},y index={3}] {./data/hdf_rtof_4_cine};
    \addplot[only marks,mark size = 1.5,green,opacity=0.4,
            mark=*,] table [x index={0},y index={1}] {./data/hdf_rtof_4_4d};
    \addplot[only marks,mark size = 1.5,blue,opacity=0.4,
            mark=*,] table [x index={0},y index={2}] {./data/hdf_rtof_4_4d};
    \addplot[only marks,mark size = 1.5,brown,opacity=0.4,
            mark=*,] table [x index={0},y index={3}] {./data/hdf_rtof_4_4d};

    \node[above,black] at (axis cs:0.82,-1.8){\scriptsize \textcolor{gray}{tof4}};                
    \end{axis}
  \end{tikzpicture}
    & 
    \hspace{-0.35cm}
  \begin{tikzpicture}
    \pgfplotsset{
        height=4cm, width=4cm,
        every tick label/.append style={font=\scriptsize}
    }
    \begin{axis}
    [
    y tick label style={
    /pgf/number format/.cd,
    precision=4,
    /tikz/.cd,
    },
    xmin=0, xmax=1, ymin = -2, ymax = 2,
    ]
    \addplot[green, solid, line width=.25pt] table [x index={0},y index={1}] {./data/hdf_rtof_5_cine};
    \addplot[blue, solid, line width=.25pt] table [x index={0},y index={2}] {./data/hdf_rtof_5_cine};
    \addplot[brown, solid, line width=.25pt] table [x index={0},y index={3}] {./data/hdf_rtof_5_cine};
    \addplot[only marks,mark size = 1.5,green,opacity=0.4,
            mark=*,] table [x index={0},y index={1}] {./data/hdf_rtof_5_4d};
    \addplot[only marks,mark size = 1.5,blue,opacity=0.4,
            mark=*,] table [x index={0},y index={2}] {./data/hdf_rtof_5_4d};
    \addplot[only marks,mark size = 1.5,brown,opacity=0.4,
            mark=*,] table [x index={0},y index={3}] {./data/hdf_rtof_5_4d};

    \node[above,black] at (axis cs:0.82,-1.8){\scriptsize \textcolor{gray}{tof5}};                
    \end{axis}
  \end{tikzpicture}
 
 \\[0cm]
  \begin{tikzpicture}
    \pgfplotsset{
        height=4cm, width=4cm,
        every tick label/.append style={font=\scriptsize}
    }
    \begin{axis}
    [
    xlabel={\footnotesize $t/T$},
    ylabel={\footnotesize $\tilde{\mathbf{F}}, \boldsymbol{f}_{\text{mom}}$  [N/L]},
    y tick label style={
    /pgf/number format/.cd,
    precision=4,
    /tikz/.cd,
    },
    xmin=0, xmax=1, ymin = -2, ymax = 2,
    ylabel absolute, ylabel style={yshift=-0.2cm},
    xlabel absolute, xlabel style={yshift=0.1cm},
    ]
    \addplot[green, solid, line width=.25pt] table [x index={0},y index={1}] {./data/hdf_normal_1_cine};
    \addplot[blue, solid, line width=.25pt] table [x index={0},y index={2}] {./data/hdf_normal_1_cine};
    \addplot[brown, solid, line width=.25pt] table [x index={0},y index={3}] {./data/hdf_normal_1_cine};
    \addplot[only marks,mark size = 1.5,green,opacity=0.4,
            mark=*,] table [x index={0},y index={1}] {./data/hdf_normal_1_4d};
    \addplot[only marks,mark size = 1.5,blue,opacity=0.4,
            mark=*,] table [x index={0},y index={2}] {./data/hdf_normal_1_4d};
    \addplot[only marks,mark size = 1.5,brown,opacity=0.4,
            mark=*,] table [x index={0},y index={3}] {./data/hdf_normal_1_4d};

    \node[above,black] at (axis cs:0.82,-1.8){\scriptsize \textcolor{gray}{n1}};   
    \end{axis}
  \end{tikzpicture}
 &
    \hspace{-0.35cm}
  \begin{tikzpicture}
    \pgfplotsset{
        height=4cm, width=4cm,
        every tick label/.append style={font=\scriptsize}
    }
    \begin{axis}
    [
    xlabel={\footnotesize $t/T$},
    y tick label style={
    /pgf/number format/.cd,
    precision=4,
    /tikz/.cd,
    },
    xmin=0, xmax=1, ymin = -2, ymax = 2.5,
    xlabel absolute, xlabel style={yshift=0.1cm},
    ]
    \addplot[green, solid, line width=.25pt] table [x index={0},y index={1}] {./data/hdf_normal_2_cine};
    \addplot[blue, solid, line width=.25pt] table [x index={0},y index={2}] {./data/hdf_normal_2_cine};
    \addplot[brown, solid, line width=.25pt] table [x index={0},y index={3}] {./data/hdf_normal_2_cine};
    \addplot[only marks,mark size = 1.5,green,opacity=0.4,
            mark=*,] table [x index={0},y index={1}] {./data/hdf_normal_2_4d};
    \addplot[only marks,mark size = 1.5,blue,opacity=0.4,
            mark=*,] table [x index={0},y index={2}] {./data/hdf_normal_2_4d};
    \addplot[only marks,mark size = 1.5,brown,opacity=0.4,
            mark=*,] table [x index={0},y index={3}] {./data/hdf_normal_2_4d};

    \node[above,black] at (axis cs:0.82,-1.8){\scriptsize \textcolor{gray}{n2}};                
    \end{axis}
  \end{tikzpicture}
   &
    \hspace{-0.35cm}
  \begin{tikzpicture}
    \pgfplotsset{
        height=4cm, width=4cm,
        every tick label/.append style={font=\scriptsize}
    }
    \begin{axis}
    [
    xlabel={\footnotesize $t/T$},
    y tick label style={
    /pgf/number format/.cd,
    precision=4,
    /tikz/.cd,
    },
    xmin=0, xmax=1, ymin = -2, ymax = 2,
    xlabel absolute, xlabel style={yshift=0.1cm},
    ]
    \addplot[green, solid, line width=.25pt] table [x index={0},y index={1}] {./data/hdf_normal_6_cine};
    \addplot[blue, solid, line width=.25pt] table [x index={0},y index={2}] {./data/hdf_normal_6_cine};
    \addplot[brown, solid, line width=.25pt] table [x index={0},y index={3}] {./data/hdf_normal_6_cine};
    \addplot[only marks,mark size = 1.5,green,opacity=0.4,
            mark=*,] table [x index={0},y index={1}] {./data/hdf_normal_6_4d};
    \addplot[only marks,mark size = 1.5,blue,opacity=0.4,
            mark=*,] table [x index={0},y index={2}] {./data/hdf_normal_6_4d};
    \addplot[only marks,mark size = 1.5,brown,opacity=0.4,
            mark=*,] table [x index={0},y index={3}] {./data/hdf_normal_6_4d};

    \node[above,black] at (axis cs:0.82,-1.8){\scriptsize \textcolor{gray}{n6}};            
    \end{axis}
  \end{tikzpicture}

   \end{tabular}
   \caption{Momentum conservation for selected normal subjects and patients. Symbols: $\tilde{\mathbf{F}}(t)$. Lines: $\mathbf{f}_\text{mom}$. Components: apex-base (green); septum-free wall (blue); diaphragm-outflow tract (brown). Flow compatibility is not enforced ($\gamma_2 = \gamma_3 = 0$). \label{fig:momentum_noFCE}}
\end{figure}
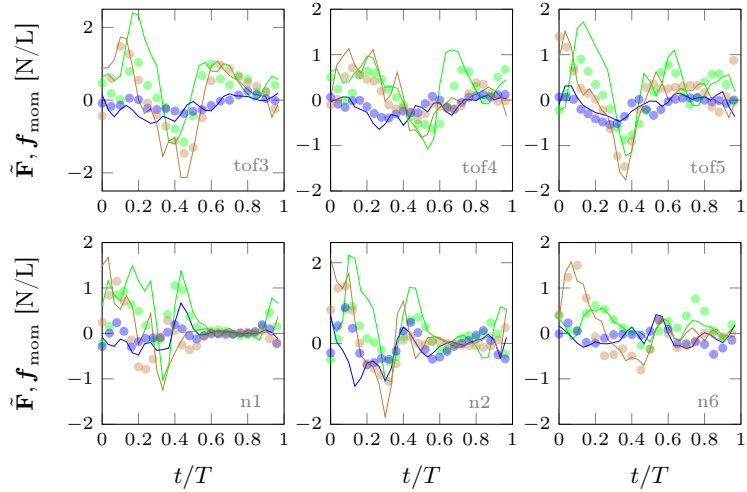

To start, the RV motion reconstruction algorithm was applied without enforcing flow compatibility, i.e., relying only on anatomical information ($\gamma_2 = \gamma_3 = 0)$, and the flow-compatible concept was used to assess the dynamic consistency of the reconstruction. \textcolor{black}{Results are reported for six selected subjects, hereinafter referred to as n1, n2 and n6 (healthy subjects), and tof3, tof4 and tof5 (patients with rTOF)}. Figure~\ref{fig:mass_rtof} shows a comparison in terms of mass conservation for the selected subjects. Overall, results are quite satisfactory when compared with very recent work~\cite{renzi2025accurate}. Nonetheless, some discrepancies can be observed: (i) the early systolic motion presents some artifacts, especially for normal subjects. These are attributed to the difficulties of the FT algorithm to follow the quick, sudden-onset motion of the RV in systole; (ii) there are under/overshoots at the systolic and/or diastolic peaks. When looking at momentum conservation, shown in Figure~\ref{fig:momentum_noFCE} for a selection of normal subjects and rTOF patients, the agreement is poorer, particularly for the normal subjects. However, medium-to-high correlation coefficients were found, indicating that intra-cohort studies are possible. For instance, the apex-base component of the hemodynamic force calculated via endocardium displacement had a cohort-averaged correlation coefficient of 0.8 with the one derived from 4D flow data, in line with previous studies \textcolor{black}{\cite{pedrizzetti2017estimating}}.

\begin{figure}
\centering \begin{tikzpicture}
    \pgfplotsset{
        height=5.2cm, width=5.2cm,
        every tick label/.append style={font=\scriptsize}
    }
    \begin{axis}
    [
    xlabel={\footnotesize $t/T$},
    ylabel={\footnotesize $\tilde{q}, f_{\text{mass}}$  [L/s]},
    y tick label style={
    /pgf/number format/.cd,
    precision=4,
    /tikz/.cd,
    },
    xmin=0, xmax=1, ymin = -0.75, ymax = 0.75,
    ylabel absolute, ylabel style={yshift=-0.2cm},
    xlabel absolute, xlabel style={yshift=0.1cm},
    legend entries={$\tilde{q}(t)$, $f_\text{mass}$, $f_\text{mass}$ (FCE)},
    legend style={fill=none, draw=none, legend pos = south east,font=\scriptsize},
    error bars/y dir=both,
    error bars/y explicit,
    ]
    \addplot[only marks,mark size = 1.5,red,opacity=0.4,
            mark=*,] table [x index={0},y index={1},y error index={2}] {./data/normal_1_open_error};
    \addplot[black,   solid, line width=.25pt] table [x index={0},y index={1}] {./data/normal_1_closed};
    \addplot[black, densely dashed, line width=.25pt] table [x index={0},y index={1}] {./data/normal_1_closed_FE_FIMH};    
    \node[above,black] at (axis cs:0.2,0.4){\scriptsize \textcolor{gray}{n1}};
    \end{axis}

  \end{tikzpicture}
  \hspace{0.7cm}
  \includegraphics[height=4.9cm]{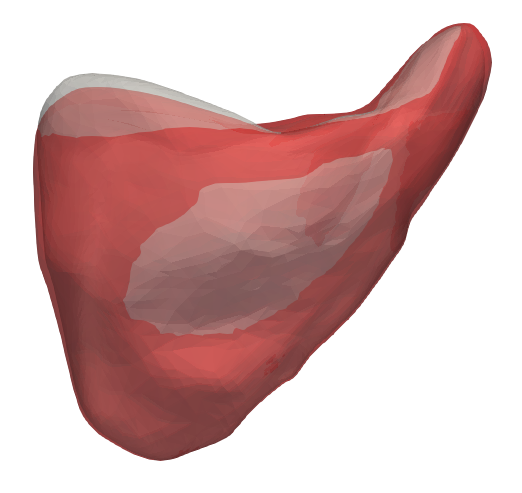}
  \caption{\textcolor{black}{(left) Mass conservation for subject n1 with and without hard enforcement of mass conservation (FCE): $\gamma_1 = 1/100; \gamma_2 = 1; \gamma_3 = 0$. (right) Comparison of RV geometry at peak systole without FCE (red surface) and with FCE (gray)}. \label{fig:FCE}}
\end{figure}
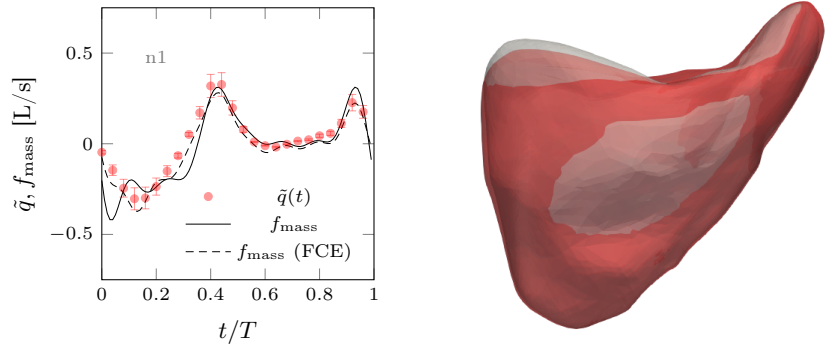

\textcolor{black}{The flow-compatibility condition was directly enforced within the reconstruction process as a correction step, as previously mentioned. In particular, initial focus has been placed on mass conservation ($\gamma_3 = 0)$. Exemplary results are shown in Fig.~\ref{fig:FCE} for normal subject n1. The enforcement of mass conservation attenuates the systolic artifact previously described and, as expected, produces a close agreement with the MRI-measured flux $\tilde{q}(t)$ (Fig.~\ref{fig:FCE}, left); of note, mass-preserving results also display an increased correlation coefficient for the hemodynamic force. Importantly, the physics-compatible enforcement resulted in only slight differences compared to the baseline reconstruction ($\gamma_2 = \gamma_3 = 0$), as shown in the right side of Fig.~\ref{fig:FCE}. At peak systole, the mean Euclidean distance and the Hausdorff distance between the two surfaces were 1.01 mm and 3.65 mm respectively; considering that the pixel size for this case was 1.41 mm, the correction is compatible with the uncertainty associated with image resolution. Additional experiments (not shown here) were also carried out with $\gamma_3 \neq 0$, resulting in better agreement with the measured hemodynamic force profiles and surface-to-surface distances in line with the values reported above.}

\section{Conclusions}

A theoretical framework for flow-compatible cardiac motion reconstruction has been introduced and tested for the challenging case of RV motion in both healthy subjects and rTOF patients, showing promising results. \textcolor{black}{When both flow and anatomical data are available, the proposed paradigm allows to integrate them to enhance the accuracy of motion reconstruction. The method is applicable to any time-resolved 3D geometry sequence, and can be extended to other chambers of the heart or to a combination of them. Future work should focus on assessing and accounting for the uncertainties associated with both anatomical and flow data, as well as on adding local (as opposed to global) constraints.}

\subsubsection{Acknowledgments}
F.C. is a Serra Húnter fellow (Generalitat de Catalunya).

\subsubsection{\discintname}
The authors have no competing interests to declare that are relevant to the content of this article. 

%
%
\bibliographystyle{splncs04}
\bibliography{Bibliography}

\begin{thebibliography}{10}
\providecommand{\url}[1]{\texttt{#1}}
\providecommand{\urlprefix}{URL }
\providecommand{\doi}[1]{https://doi.org/#1}

\bibitem{beg2005computing}
Beg, M.F., Miller, M.I., Trouv{\'e}, A., Younes, L.: Computing large
  deformation metric mappings via geodesic flows of diffeomorphisms.
  International Journal of Computer Vision  \textbf{61}(2),  139--157 (2005)

\bibitem{chen2020deep}
Chen, C., Qin, C., Qiu, H., Tarroni, G., Duan, J., Bai, W., Rueckert, D.: Deep
  learning for cardiac image segmentation: a review. Frontiers in
  cardiovascular medicine  \textbf{7}, ~25 (2020)

\bibitem{claus2015tissue}
Claus, P., Omar, A.M.S., Pedrizzetti, G., Sengupta, P.P., Nagel, E.: Tissue
  tracking technology for assessing cardiac mechanics: principles, normal
  values, and clinical applications. JACC: Cardiovascular Imaging
  \textbf{8}(12),  1444--1460 (2015)

\bibitem{durrleman2014morphometry}
Durrleman, S., Prastawa, M., Charon, N., Korenberg, J.R., Joshi, S., Gerig, G.,
  Trouv{\'e}, A.: Morphometry of anatomical shape complexes with dense
  deformations and sparse parameters. NeuroImage  \textbf{101},  35--49 (2014)

\bibitem{hansen2014method}
Hansen, M.S., Olivieri, L.J., O'Brien, K., Cross, R.R., Inati, S.J., Kellman,
  P.: Method for calculating confidence intervals for phase contrast flow
  measurements. Journal of Cardiovascular Magnetic Resonance  \textbf{16}(1),
  ~46 (2014)

\bibitem{kovacs2019right}
Kov{\'a}cs, A., Lakatos, B., Tokodi, M., Merkely, B.: Right ventricular
  mechanical pattern in health and disease: beyond longitudinal shortening.
  Heart Failure Reviews  \textbf{24}(4),  511--520 (2019)

\bibitem{menon2024cardiovascular}
Menon, K., Hu, Z., Marsden, A.L.: Cardiovascular fluid dynamics: a journey
  through our circulation. Flow  \textbf{4}, ~E7 (2024)

\bibitem{mittal2016computational}
Mittal, R., Seo, J.H., Vedula, V., Choi, Y.J., Liu, H., Huang, H.H., Jain, S.,
  Younes, L., Abraham, T., George, R.T.: Computational modeling of cardiac
  hemodynamics: current status and future outlook. Journal of Computational
  Physics  \textbf{305},  1065--1082 (2016)

\bibitem{morales2019implementation}
Morales, M.A., Izquierdo-Garcia, D., Aganj, I., Kalpathy-Cramer, J., Rosen,
  B.R., Catana, C.: Implementation and validation of a three-dimensional
  cardiac motion estimation network. Radiology: Artificial Intelligence
  \textbf{1}(4),  e180080 (2019)

\bibitem{pedrizzetti2019computation}
Pedrizzetti, G.: On the computation of hemodynamic forces in the heart
  chambers. Journal of Biomechanics  \textbf{95},  109323 (2019)

\bibitem{pedrizzetti2017estimating}
Pedrizzetti, G., Arvidsson, P.M., T{\"o}ger, J., Borgquist, R., Domenichini,
  F., Arheden, H., Heiberg, E.: On estimating intraventricular hemodynamic
  forces from endocardial dynamics: a comparative study with 4d flow mri.
  Journal of biomechanics  \textbf{60},  203--210 (2017)

\bibitem{renzi2025accurate}
Renzi, F., Vergara, C., Fedele, M., Giambruno, V., Quarteroni, A., Puppini, G.,
  Luciani, G.B.: Accurate reconstruction of right heart shape and motion from
  cine-mri for image-driven computational hemodynamics. International Journal
  for Numerical Methods in Biomedical Engineering  \textbf{41}(1),  e3891
  (2025)

\bibitem{soulat20204d}
Soulat, G., McCarthy, P., Markl, M.: 4d flow with mri. Annual review of
  biomedical engineering  \textbf{22}(1),  103--126 (2020)

\bibitem{voigt20192}
Voigt, J.U., Cvijic, M.: 2-and 3-dimensional myocardial strain in cardiac
  health and disease. JACC: Cardiovascular Imaging  \textbf{12}(9),  1849--1863
  (2019)

\end{thebibliography}

\end{document}